\documentclass[a4paper,12pt]{article}
\usepackage[english]{babel}
\usepackage{setspace}
\usepackage[T1]{fontenc}
\usepackage{amsfonts}
\usepackage[verbose,bookmarksopen=false,hypertexnames=false,linktocpage,bookmarksnumbered,colorlinks,urlcolor=blue,citecolor=blue]{hyperref}
\usepackage{amsmath}
\usepackage{amssymb}
\usepackage{amsthm}
\usepackage{listings}
\usepackage{float}
\usepackage{graphicx}
\usepackage[font=small,format=hang]{caption}
\usepackage{epsfig}
\usepackage{mathrsfs}
\usepackage{fancyhdr}
\usepackage{hhline}
\usepackage{calc}
\usepackage{url}
\usepackage{natbib}
\makeatletter
\def\iffloatpage#1#2{\if@fcolmade#1\else#2\fi}
\makeatother
\usepackage{fullpage}
\usepackage{colortbl}
\usepackage{alltt}
\usepackage{color}
	\definecolor{string}{rgb}{0.7,0.0,0.0}
	\definecolor{comment}{rgb}{0.13,0.54,0.13}
	\definecolor{keyword}{rgb}{0.0,0.0,1.0}
	\definecolor{gris25}{gray}{0.75}
\usepackage{enumerate}
\usepackage{lscape}
\def\frac#1#2{{\textstyle{#1\over#2}}}

\DeclareSymbolFont{AMSb}{U}{msb}{m}{n}
\DeclareMathSymbol{\Natural}{\mathbin}{AMSb}{"4E}
\DeclareMathSymbol{\Integer}{\mathbin}{AMSb}{"5A}
\DeclareMathSymbol{\Real}{\mathbin}{AMSb}{"52}
\DeclareMathSymbol{\Rational}{\mathbin}{AMSb}{"51}
\DeclareMathSymbol{\Imaginary}{\mathbin}{AMSb}{"49}
\DeclareMathSymbol{\Complex}{\mathbin}{AMSb}{"43} 
\DeclareMathSymbol{\Disk}{\mathbin}{AMSb}{"44} 
\def\bi{\begin{itemize}}
\def\ei{\end{itemize}}
\def\bd{\begin{description}}
\def\ed{\end{description}}
\def\ben{\begin{enumerate}}
\def\een{\end{enumerate}}
\def\bv{\begin{verbatim}}
\def\ev{\end{verbatim}}

\def\cth{{$c{\rm th}$ }}

\def\rth{{$r{\rm th}$ }}

\def\calB{{\mathcal B}}

\def\calD{{\mathcal D}}

\def\calK{{\mathcal K}}

\def\calN{{\mathcal N}}

\def\calS{{\mathcal{S}}}

\def\calT{{\mathcal T}}

\def\calX{{\mathcal X}}

\def\v{{\varepsilon}}

\def\hat#1{{\widehat{#1}}}

\def\pr{{\rm Pr}}

\def\E{{\rm E}}

\def\var{{\rm var}}
\def\avar{{\rm avar}}

\def\Dto{{\ {\buildrel D\over \longrightarrow}\ }}

\def\2to{{\ {\buildrel 2\over \longrightarrow}\ }}

\def\iid{{\ {\buildrel \rm{iid}\over \sim}\ }}

\def\Deq{{\ {\buildrel D\over =}\ }}

\def\I1ton{{$I_1,\ldots,I_n$}}
\def\X1ton{{$X_1,\ldots,X_n$}}
\def\Y1ton{{$Y_1,\ldots,Y_n$}}
\def\Z1ton{{$Z_1,\ldots,Z_n$}}
\def\R1ton{{$R_1,\ldots,R_n$}}
\def\e1ton{{$e_1,\ldots,e_n$}}
\def\t1ton{{$t_1,\ldots,t_n$}}
\def\x1ton{{$x_1,\ldots,x_n$}}
\def\y1ton{{$y_1,\ldots,y_n$}}
\def\z1ton{{$z_1,\ldots,z_n$}}

\def\calS{{\mathcal{S}}}

\newtheorem{defn}{Definition}

\newtheorem{theorem}[defn]{Theorem}

\title{Space-time modelling of extreme events}

\author{Rapha\"el Huser\thanks{Ecole Polytechnique F\'ed\'erale de Lausanne, EPFL-FSB-MATHAA-STAT, Station 8, 1015 Lausanne, Switzerland. } \, and\, A.~C.~Davison$^*$}

\begin{document}

\renewcommand{\textfraction}{0.05}
\renewcommand{\topfraction}{0.95}
\renewcommand{\bottomfraction}{0.95}
\renewcommand{\floatpagefraction}{0.35}

\onehalfspacing
\pagenumbering{roman}
\setcounter{page}{0}\pagenumbering{arabic}

\maketitle

\subsection*{Abstract}
Max-stable processes are the natural analogues of the generalized extreme-value distribution for the modelling of extreme events in space and time.  Under suitable conditions, these processes are asymptotically justified models for maxima of independent replications of random fields, and they are also suitable for the modelling of joint individual extreme measurements over high thresholds. This paper extends a model of \citet{Schlather01} to the space-time framework, and shows how a pairwise censored likelihood can be used for consistent estimation under mild mixing conditions. Estimator efficiency is also assessed and the choice of pairs to be included in the pairwise likelihood is discussed based on computations for simple time series models. The ideas are illustrated by an application to hourly precipitation data over Switzerland.
\vskip 1in

\noindent\textbf{Keywords:} Composite likelihood; Extremal coefficient; Max-stable process; Rainfall data; Random set; Threshold-based inference.

\newpage

\section{Introduction}
Under suitable conditions, max-stable processes are asymptotically justified models for maxima of independent replications of random fields.  Since they extend the generalized extreme-value distribution of univariate extreme value theory to the functional setting, they thus appear to be natural models for spatial extremes. \citeauthor{deHaan01}'s~(1984) spectral representation of such processes implies that there are infinitely many max-stable processes, and in practice the challenge is to build flexible but parsimonious models that can capture a wide range of extremal dependencies. Parsimony is important since extremal data are often scarce, but  flexibility is also crucial since a poor fit might lead to mis-estimation of the risk. Several models for max-stable processes have been proposed: \citet{Smith01} proposes a max-stable model with deterministic storm shapes, and \citet{Schlather01} proposes a model based on a Gaussian process. Other models include the Brown--Resnick processes \citep[see][]{Kabluchko01}, or a Brownian motion model proposed by \citet{Buishand01}, which has the drawback of not being invariant with respect to coordinate axes. \citet{Wadsworth01} generalize these max-stable models to hybrid spatial dependence models able to capture and handle both asymptotic dependence and asymptotic independence. \citet{Reich01} propose a finite-dimensional construction of max-stable processes that can be fitted in the Bayesian framework with Markov chain Monte Carlo methods. Other modelling approaches for spatial extremes, based either on copula or on latent processes, are presented by \citet{Davison06}. 

The full likelihood cannot be obtained analytically for most max-stable processes \citep[but see][]{Genton03}. However, since the bivariate marginal densities can usually be derived, inference can be based on a composite likelihood. Much has been written on pseudo-, quasi- or composite-likelihood: see for example \citet{Hjort01}, \citet{Lindsay01}
, \citet{Varin01}, \citet{Varin02}, \citet{Cox01}, or \citet{Padoan01} and \citet{Davison03} for applications to spatial extremes.   Such likelihoods are robust to misspecification of higher distributional assumptions and have nice theoretical properties, but so far have been applied only to componentwise maxima. An important extension, which improves inference by incorporating more information, is to perform pairwise threshold-based inference for max-stable processes, analogous to the use of the generalized Pareto distribution in the univariate case. This will be addressed in this article.


In Section~\ref{ModellingSection}, we tie together geostatistics and statistics of extremes to construct asymptotically valid space-time models for extremes. The spatio-temporal aspect of this modelling is novel, though related work include \citet{Davis01} and \citet{Davis04}. Section~\ref{InferenceSection} is focused on inference and describes the methods based on pairwise likelihood, while Section~\ref{EfficiencySection} addresses the loss in efficiency of the estimation procedure compared to classical maximum likelihood estimation and gives some suggestions about the choice of pairs to be included in the pairwise likelihood. Section~\ref{SimulationSection} describes simulations to validate our approach, and Section~\ref{ApplicationSection} describes its application to space-time modelling of rainfall. Some concluding discussion is given in Section~\ref{DiscussionSection}.

\section{Threshold modelling for extremes}\label{ModellingSection}
\subsection{Marginal modelling}\label{UnivariateEVT}

The classical theory of extreme values addresses the large-sample fluctuations of the maximum $M_n$ of a sequence of independent and identically distributed random variables $X_1,\ldots,X_n$ whose distribution $F$ has upper terminal $x_F=\sup\{x:F(x)<1\}\in\Real\cup\{+\infty\}$. If sequences $\{a_n\}>0$ and $\{b_n\}\subset\Real$ exist such that $(M_n-b_n)/a_n$ converges in distribution to a non-degenerate distribution $G$, then this must necessarily be the generalized extreme-value (GEV) distribution, that is $G(x)=\exp[-\{1+\xi(x-\eta)/\tau\}^{-1/\xi}]$, defined on the set $1+\xi(x-\eta)/\tau>0$, with $\eta\in\Real,\tau>0,\xi\in\Real$ and with the value for $\xi=0$ being interpreted as $\xi\to 0$. A complementary result describes the stochastic behaviour of peaks over a high threshold $u$: if this limiting result holds for maxima, then as $u\to x_F$ the conditional distribution of $X-u$, given that $X>u$, converges to the generalized Pareto distribution,  GPD$(\sigma,\xi)$ \citep{Davison04}. The distribution of such a variate is
$$
H(y) = 1 - \left(1 + {\xi\over \sigma}y\right)^{-1/\xi},\qquad y>0,
$$
where the scale parameter is linked to that of the GEV distribution by $\sigma=\tau + \xi(u-\eta)$. A closely related characterization of extremes relies on point processes. If the limiting result holds for maxima, then as $n\to\infty$ the two-dimensional point process $\{i/(n+1),(X_i-b_n)/a_n\}_{i=1}^n$ converges to a non-homogenous Poisson process on regions of the form $[t_1,t_2]\times[u,\infty)$, ${0<t_1<t_2<1}$, with a certain intensity \citep[see][]{Leadbetter03,Smith02}. In practice, the data often exhibit temporal dependence, and the aforementioned asymptotic results can be extended to stationary sequences with short-range dependence \citep[see][]{Leadbetter01}, where serial dependence of the extremes is summarized by the extremal index. For more details about extreme-value statistics, see \citet{Coles01}, \citet{Beirlant01}, \citet{Embrechts01} or \citet{DeHaan02}. As the tail may be well approximated by a GPD, the distribution $F$ of $X$ can be consistently estimated by
$$
\tilde F(x) = \left\{\begin{array}{ll}
\hat F(x), & x\leq u; \\
1 - \hat\zeta_u \left\{1 + {\hat\xi\over \hat\sigma}(x-u)\right\}^{-1/\hat\xi}, & x>u,
\end{array}\right.
$$
where $\hat F(x)$ is the empirical distribution function of the sample $X_1,\ldots,X_n$, $\hat\zeta_u$ is the estimated probability of exceeding the threshold $u$ and $\hat\xi$ and $\hat\sigma$ are estimates of $\xi$ and $\sigma$. The transformation $t(x)=-1/\log \tilde F(x)$ therefore approximately standardizes the observations to have the unit Fr\'echet distribution $\exp(-1/x)$ for $x>0$.

Joint modelling of extremes is crucial for a realistic assessment of risk, and the next section describes models for spatial or spatio-temporal extremes, where the margins have been previously transformed to the unit Fr\'echet scale. 

\subsection{Max-stable processes}\label{MSproc}
A spatial random process $Z(x)$ defined for $x\in\calX\subset\Real^d$ and with unit Fr\'echet margins is said to be \textit{max-stable} if for any finite set $\calD\subset\calX$ and any function $z(x)$ defined on $\calD$, the following property is satisfied:
$$
\pr\left\{{Z(x)/ n}\leq z(x), x \in\calD\right\}^n = \pr\left\{Z(x)\leq z(x), x \in\calD\right\},\qquad n=1,2,\ldots.
$$
As the class of GEV distributions coincides with that of univariate max-stable distributions, the marginal distributions of a max-stable process must be GEV, and therefore, if $\{Y_i(x):x\in\calX\subset\Real^d\}$, $i=1,2,\ldots$, are independent and identically distributed replications of a random process with arbitrary margins, and if there exist sequences of continuous functions $\{a_n(x)\}>0$ and $\{b_n(x)\}$ such that 
$$
a_n^{-1}\{\max(Y_1,\ldots,Y_n)-b_n\}\Dto Z^*,\quad n\to\infty, 
$$
where $Z^*$ is a non-degenerate random field, then $Z^*$ must be max-stable with GEV margins. Consequently, the only possible non-degenerate limits for properly linearly renormalized maxima of random processes are max-stable processes, which are therefore asymptotically justified models for spatial extremes. \citet{deHaan01} proved that a process $Z$ with unit Fr\'echet margins is max-stable if and only if it can be represented as
\begin{equation}
\label{spectralRepresentation}
Z(x)=\sup_{i\geq 1}\xi_i W_{i}(x),
\end{equation}
where the $\xi_i$'s are the points of a Poisson process on $\Real_+$ with intensity $\xi^{-2}d\xi$ and where the $W_{i}$'s are independent replicates of a non-negative random process $W(x)$ with mean $1$. We can think of the $W_i$'s as random storms in space and of the $\xi_i$'s as their intensities. Due to the characterization (\ref{spectralRepresentation}), no finite parametrization exists for max-stable processes. 

From (\ref{spectralRepresentation}), it can be straightforwardly shown that the joint distribution of the process $Z$ at $N$ distinct locations is
\begin{equation}
\label{ModelForMaxima}
\pr\{Z(x_1)\leq z_1,\ldots,Z(x_N)\leq z_N\}=\exp\left(-E\left[\max_{i=1,\ldots,N}\left\{W(x_i)\over z_i\right\}\right] \right)=\exp\{-V_N(z_1,\ldots,z_N)\},
\end{equation}
where the exponent measure $V_N(\cdot)$, which summarises the extremal dependence structure, is homogeneous of order $-1$ and satisfies $V_N(\infty,\ldots,z,\ldots,\infty)=1/z$ for any permutation of the $N$ arguments.  When $z_i=z$ for all $i=1,\ldots,N$, we obtain $\pr\{Z(x_1)\leq z,\ldots,Z(x_N)\leq z\}=\exp\{-V_N(1,\ldots,1)/z\}=\{\exp(-1/z)\}^{V_N(1,\ldots,1)}$, so the so-called extremal coefficient $\theta_N=V_N(1,\ldots,1)$ can be seen as a summary of extremal dependence, and has two bounding cases: complete dependence, $\theta_N=1$, and asymptotic independence, $\theta_N=N$.

Different choices for $W(x)$ yield different models for spatial maxima, with more or less flexible dependence structures. For our purpose, i.e.,\ the modelling of extreme rainfall (see \S\ref{ApplicationSection}), the model proposed by \citet{Davison03} and originally due to \citet{Schlather01}, seems suitable. They consider a truncated Gaussian random process for $W(x)$, so that storm shapes are stochastic, and include a compact random set, which allows one to capture complete independence of the extremes. The model is defined by taking
\begin{equation}
\label{SchlatherModel}
W_i(x)\propto\max\{0,\v_i(x)\}I_{\calB_i}(x-X_i),\qquad x\in\calX,
\end{equation}
where $\calX$ is compact in $\Real^d$ and the $\v_i$ are independent replicates of a Gaussian random field with correlation function $\rho(h)$, $I_\calB$ is the indicator function of a compact random set $\calB\subset\calX$,  the $\calB_i$ are independent replications of $\calB$, and the $X_i$ are points of a Poisson process of unit rate on $\calX$, independent of the $\v_i$. The proportionality constant in (\ref{SchlatherModel}) is chosen to satisfy $\E\{W_i(x)\}=1$.

A common feature of the max-stable models thus far proposed is that the exponent measure $V_N$ is known for $N=2$. \citet{Genton03} provide a closed-form expression of the likelihood function for the Smith max-stable model indexed by $\Real^d$ at $N\leq d+1$ sites  ($d\geq 1$), but typically only the bivariate margins are known. Moreover the number of terms involved in the likelihood increases at a combinatorial rate as $N$ increases. Therefore, standard likelihood-based inference seems to be out of reach. Following \citet{Davison03}, a pairwise likelihood approach is considered (see Section~\ref{InferenceSection}). The bivariate exponent measure for the model with (\ref{SchlatherModel}) can be expressed in the stationary case as
\begin{equation}\label{ExponentMeasureSchlatherModel}V_2(z_1,z_2)=\left({1\over z_1} + {1\over z_2}\right)\left\{1-{\alpha(h)\over 2}\left(1-\left[1-2{\{\rho(h)+1\}z_1z_2\over (z_1+z_2)^2}\right]^{1/2}\right)\right\},\end{equation}
where $h=x_1-x_2$, $\alpha(h)=\E\{|\calB\cap(h+\calB)|\}/\E(|\calB|)$ and $|\cdot|$ is used to denote the volume of a set. Hence the pairwise extremal coefficients are 
\begin{equation}\label{ExtrCoeffDefinition}
\theta_2(h)=V_2(1,1)=2-\alpha(h)\left\{1-\sqrt{{1-\rho(h)\over 2}}\right\}.
\end{equation}
As mentioned in \citet[][p.38]{Abrahamsen01}, a valid isotropic correlation function $\rho(h)$ in $\Real^2$ satisfies $\rho(h)>-0.403$. Therefore, if there were no random set $\calB$, i.e.,\  $\calB\equiv\calX$ and $\alpha(h)\equiv1$, $\theta_2(h)$ would be bounded above by $1.838$ and complete independence could not be captured by this model even at very large distances. With this model, and since $\calB$ is chosen to be compact, for modelling purposes we can choose $\calB$ so that $\alpha(h)\to 0$ and thus $\theta_2(h)\to 2$ as $h\to\infty$ for any correlation function $\rho(h)$. This model is built from random sets with a Schlather model inside, so the short-range dependence is largely determined by the correlation $\rho(h)$, while the longer-range dependence is regulated by the geometry of the random set $\calB$.  There are clearly other possibilities for the model inside the random set, but for concreteness we consider just one here. 

In a more general framework, the correlation function need be neither isotropic nor stationary, and could therefore depend on the spatial locations $x_1$ and $x_2$ rather than on their distance $\|h\|$ and lag vector $h$. We would then have non-stationary extremal coefficients. 

In the context of modelling space-time extremes the points $x\in\calX$ have coordinates in space $\calS=\Real^2$ and time $\calT\in\Real$, that is, $x=(s,t)\in\calX=\calS\times\calT$. The function $\rho$ must therefore be a valid space-time correlation function \citep{Gneiting01,Cressie02,Davis04}.

In the following section, we show how to make the link from the asymptotic distribution for maxima to a joint model for the right tail.

\subsection{Censored threshold-based likelihood}
The convergence of block maxima to a max-stable process implies that all finite-dimensional distributions converge to a max-stable distribution, i.e., to  a multivariate extreme value distribution. Let $\{Y_n(x):x\in\calX\subset\Real^d\}$, $n=1,2,\ldots$ be independent and identically distributed replicates of a process $Y(x)$ with Fr\'echet margins. As explained in Section~\ref{MSproc}, the joint distribution of properly scaled block maxima at $N$ sites in $\calX$ is well approximated by $\exp\{-V_N(z_1,\ldots,z_N)\}$, where the exponent measure $V_N$ stems from the underlying spatial structure of the max-stable process. Hence, for a large fixed $n$, the joint distribution at $N$ locations is
\begin{eqnarray}
\pr\{Y(x_1)\leq z_1,\ldots, Y(x_N)\leq z_N\}& = & \left(\left[\pr\{Y(x_1)\leq z_1,\ldots, Y(x_N)\leq z_N\}\right]^n\right)^{1/n}\nonumber \\
& = & \pr\left\{\max_{i=1,\ldots,n}Y_i(x_1)\leq z_1,\ldots, \max_{i=1,\ldots,n}Y_i(x_N)\leq z_N\right\}^{1/n}\nonumber\\
& \approx & \exp\left\{-V_N\left({z_1\over n},\ldots,{z_N\over n}\right)\right\}^{1/n}\nonumber\\
& = & \exp\left\{-{1\over n}V_N\left({z_1\over n},\ldots,{z_N\over n}\right)\right\}\nonumber\\
& = & \exp\left\{-V_N\left({z_1},\ldots,{z_N}\right)\right\}\label{distr},
\end{eqnarray}
the last equality coming from the homogeneity of $V_N$. Hence, the model for maxima in equation~(\ref{ModelForMaxima}) also provides a model for rare events of individual observations. This approximation is only good for large positive values $z_1,\ldots,z_N\in\Real$, since the impact of the approximation is negligible when $\pr\{Y(x_1)\leq z_1,\ldots, Y(x_N)\leq z_N\}$ is close to $1$. Hence, the bivariate joint density of the process $Y$ at locations $x_1$ and $x_2$ has the form $\partial^2 \exp\left\{-V_2\left({z_1},{z_2}\right)\right\} / \partial z_1\partial z_2$ for large $z_1,z_2$. However, as this model is only valid when the two events are simultaneously extreme at both locations, we adopt a censored likelihood approach \citep[see][p.155]{Coles01}. Let the threshold $u$ be sufficiently high that equation~(\ref{distr}) is a valid model for $z_1,z_2>u$. Then the likelihood contribution $p_u(z_1,z_2)$ of a pair $(z_1,z_2)$ is
\begin{eqnarray}\label{censoredLik}
p_u(z_1,z_2)=\left\{\begin{array}{ll}
{\partial^2\over\partial z_1\partial z_2} \exp\{-V_2(z_1,z_2)\}, & z_1,z_2>u;\\
{\partial\over\partial z_1} \exp\{-V_2(z_1,u)\}, & z_1>u,z_2\leq u;\\
{\partial\over\partial z_2} \exp\{-V_2(u,z_2)\}, & z_1\leq u,z_2> u;\\
\exp\{-V_2(u,u)\}, & z_1,z_2\leq u.
\end{array}\right.
\end{eqnarray}
Different marginal thresholds could be used \citep{Bortot02} and the approach could be generalized to higher dimensions. However, in practice, the probability that an observed $N$-uplet falls into the ``upper right quadrant'' decays geometrically with $N$, leading to inference problems. In the next section, we will show that these censored threshold-based pairwise likelihood contributions provide consistent inference.

\section{Inference}\label{InferenceSection}
\subsection{Pairwise likelihood approach}

As the full likelihood is not known for max-stable models, classical frequentist or Bayesian inference is impossible, and we adopt an alternative approach based on composite likelihood. An analogous approach in the Bayesian framework using a pseudo-posterior distribution based on a pairwise likelihood has been developed by \citet{Ribatet02}. Maximum composite likelihood estimators typically have similar asymptotic properties to the usual maximum likelihood estimator; often they are asymptotically normal and strongly consistent.

Assume that the spatio-temporal process $Z(x)$, $x=(s,t)\in\calX=\calS\times\calT$ is observed at $S$ monitoring stations and at times $1,\ldots,T$, that is at $N=ST$ locations in $\calX$. Let $z_{s,t}$ denote the observation recorded at station $s$ and time $t$, and consider the censored threshold-based pairwise likelihood
\begin{equation}
\label{pairwiselik}
l_\calK(\theta)=\sum_{t=1}^T\sum_{h\in\calK_t}\sum_{s_1=1}^S\sum_{s_2=1}^S (1-I\{s_1\geq s_2 \mbox{ and } h=0\})\log p_u\left(z_{s_1,t},z_{s_2,t+h};\theta\right),
\end{equation}
with the corresponding maximum pairwise likelihood estimator
\begin{equation}
\label{pairwiselikEst}\hat\theta_{p,\calK} = \arg\max_{\theta\in\Theta} l_\calK(\theta),
\end{equation}
where $\calK_t=\{h\in\calK: h\leq T-t\}$ and $\calK\subset\Natural\cup\{0\}$ is a finite collection of time lags, where $p_u$ is given by equation~(\ref{censoredLik}), the exponent measure $V$ being given for example by (\ref{ExponentMeasureSchlatherModel}) and where $I\{\cdot\}$ is the indicator function. If $\calK=\{0,1,\ldots,K\}$ for $K<\infty$, this pairwise likelihood corresponds to summing up all space-time pairwise contributions, up to a maximum time lag $K$. If $K=T-1$, it reduces to the full pairwise likelihood. However, the associated computational burden could be reduced and statistical efficiency gained by taking a different subset $\calK$. For example, we could take $\{\lfloor a^{k-1}\rfloor:k=1,\ldots,K\}\cup\{0\}$, $a>1$. In particular, when $a=2$, we include the pairs at lag $0,1,2,4,8,\ldots$. Another choice could be based on the Fibonacci sequence: $0,1,2,3,5,8,13,\ldots$. In Section~\ref{EfficiencySection}, we will see that the choice of pairs is closely linked to the efficiency of $\hat\theta_{p,\calK}$ and thus, a careful selection of them is essential.

\subsection{Asymptotics}
\citet{Davison03} and \citet{Padoan01} use pairwise likelihood for inference on max-stable processes, assuming independence between distinct annual maxima. In the case of spatio-temporal extremes, the asymptotic normality of $\hat\theta_{p,\calK}$ stems from a central limit theorem for stationary time series applied to the score $U(\theta)=\nabla l(\theta)=\sum_{t=1}^T U_t(\theta)$, where $U_t(\theta)$ is the derivative of rightmost triple sums in equation~(\ref{pairwiselik}) with respect to $\theta$. However, as the elements $U_t(\theta)$ are generally correlated over time $t$, we need an additional mixing condition in order for classical asymptotics to hold. A suitable mild sufficient condition is that the process $Z(x)$ be temporally \textit{$\alpha$-mixing}, along with a condition on the rate at which the mixing coefficients $\alpha(n)$ must decay, ensuring that the correlation vanishes sufficiently fast at infinity. With this condition, two events become more and more independent as their time lag increases. In particular, all $m$-dependent processes are contained within the class of $\alpha$-mixing processes.

We call a space-time process $Z(x),x=(s,t)\in\calX=\calS\times\calT$ temporally $\alpha$-mixing with coefficients $\alpha(n)$ if for all $s\in\calS$, for all sequences $t_n\subset\calT$, the time series $\{Z(s,t_n),n\in\Natural\}$ is $\alpha$-mixing with coefficients $\alpha_s(n)$ and where $\sup_{s\in\calS}\alpha_s(n)\leq\alpha(n)\to 0$ as $n\to\infty$. For the definition of an $\alpha$-mixing time series, see \citet[Definition 1.6]{Bradley01}. We can then obtain the following theorem, whose proof, which relies on the theory of estimating equations, is given in Appendix \ref{ProofOfThm1}.

\begin{theorem}
Assume that $Z(x)$ is a stationary spatio-temporal max-stable process which is temporally $\alpha$-mixing with coefficients $\alpha(n)$. Moreover, suppose that for all $\theta\in\Theta$, $\E[\{U_1(\theta)\}^2]<\infty$ and that for some $\delta>0$, one has ${\E(|U_1(\theta)|^{2+\delta})<\infty}$ and ${\sum_{n\geq 1}|\alpha(n)|^{\delta/(2+\delta)}<\infty}$. Then, if $\theta$ is identifiable from the bivariate densities, then 
$$
T^{1/2}K(\theta)^{-1/2}J_1(\theta)(\hat\theta_{p,\calK}-\theta) \Dto \calN\left(0,I_p\right),
$$
where
\begin{eqnarray}
J_1(\theta) & = & \E\{-\nabla_\theta U_1(\theta)\}; \label{Bread}\\
K(\theta) & = & T^{-1}\var\left\{\sum_{t=1}^TU_t(\theta)\right\} \nonumber\\
& = & \E\{U_1(\theta)U_1(\theta)^T\} + \sum_{t=1}^{T-1}\left(1-{t\over T}\right)\left[\E\left\{U_1(\theta)U_{t+1}(\theta)^T\right\} + \E\left\{U_{t+1}(\theta)U_t(\theta)^T\right\}\right]\label{Meat}\\
&\to & \E\{U_1(\theta)U_1(\theta)^T\} + \sum_{t=1}^{\infty}\left[\E\left\{U_1(\theta)U_{t+1}(\theta)^T\right\} + \E\left\{U_{t+1}(\theta)U_t(\theta)^T\right\}\right]<\infty,\qquad  T\to\infty.\nonumber
\end{eqnarray}
\end{theorem}

This result shows that the standard asymptotic normality result for composite likelihoods \citep{Hjort01,Lindsay01,Godambe01,Varin01,Varin02,Cox01,Padoan01} still holds under mild conditions for moderately temporally dependent processes. Furthermore, the asymptotic variance turns out to be of ``sandwich'' form, as is standard for misspecified models. 

If the process $Z(x)$ were instead assumed to be Gaussian, and hence not max-stable, and if the pairwise likelihood were defined in terms of the marginal bivariate normal densities, then the moment conditions of the theorem, i.e., $\E[\{U_1(\theta)\}^2]<\infty$ and ${\E(|U_1(\theta)|^{2+\delta})<\infty}$, would be automatically satisfied for all $\delta>0$, and thus the mixing condition would reduce to $\sum_{n\geq 1}|\alpha(n)|^{1-\epsilon}<\infty$, for some $\epsilon>0$. Similar results were obtained by \citet{Davis03}, who establish the asymptotic normality and the strong consistency of the maximum consecutive pairwise likelihood estimator for ARMA models, under a condition on the autocorrelation function. They also treat certain long-memory models.

\subsection{Variance estimation}\label{VarianceEstimation}
Variance estimation for $\hat\theta_{p,\calK}$ is difficult due to the complicated form of the sandwich matrices in equations (\ref{Bread}) and (\ref{Meat}). The pairwise log likelihood is formed by summing the pairwise contributions for the time lags in the set $\calK$ and across all $S$ stations, so a single evaluation of the pairwise log likelihood requires O$(T|\calK| S^2)$ operations, and the computation of (\ref{Meat}) is still more intensive.

The temporal dependence of the data suggests that block bootstrap or jackknife methods be used. For computational reasons, in our application we choose to apply a block jackknife, treating rainfall data from different summers as independent. For that purpose, we leave out yearly blocks one at a time, and get pseudo-values of $\hat\theta_{p,\calK}$ to estimate its variability, using the formula of \citet{Busing01}. Fortunately, the pseudo-values can be computed in parallel.

\section{Efficiency considerations}\label{EfficiencySection}
In Section~\ref{InferenceSection}, we introduced our maximum pairwise likelihood estimator for spatio-temporal extremes. Although it inherits its asymptotic properties from the traditional maximum likelihood estimator, the natural question of statistical efficiency remains to be addressed. It turns out that the loss in efficiency is closely related to the pairs that are included in the pairwise likelihood, that is to the choice of $\calK$. Adding pairs might simultaneously increase the variability $K(\theta)$ of the score and the amount of information, $J(\theta)$, so it is unclear how the selection of pairs acts on the variance $T^{-1}J(\theta)^{-1}K(\theta)J(\theta)^{-1}$; the amount of information contained in a single pair might be insufficient to counteract the increase of variability due to including it, so the choice of the optimal subset of pairs is not obvious. However, one might suspect that for short-range dependent processes, the pairs that are far apart in $\calS\times\calT$ are not as relevant for the estimation of a dependence parameter as are the close ones. \citet{Varin04}, \citet{Varin02} and \citet{Varin05} already suggested the elimination of non-neighbouring pairs.

We studied the efficiency of the maximum pairwise likelihood estimator for time series models whose maximum likelihood estimators could be computed, hoping to gain a qualitative understanding of how composite likelihoods behave in more complex settings. In the same vein as \citet{Davis03}, we studied AR$(1)$ and MA$(1)$ processes, but with the different objective of understanding how the asymptotic relative efficiency evolves as the set of time lags $\calK$ for the selection of pairs in the likelihood varies. Complementary results on the efficiency of pairwise likelihood may be found in \citet{Cox01}, \citet{Varin03}, \citet{Hjort01}  or \citet{Joe01}.

\begin{figure}
\centering
\includegraphics[width=16cm]{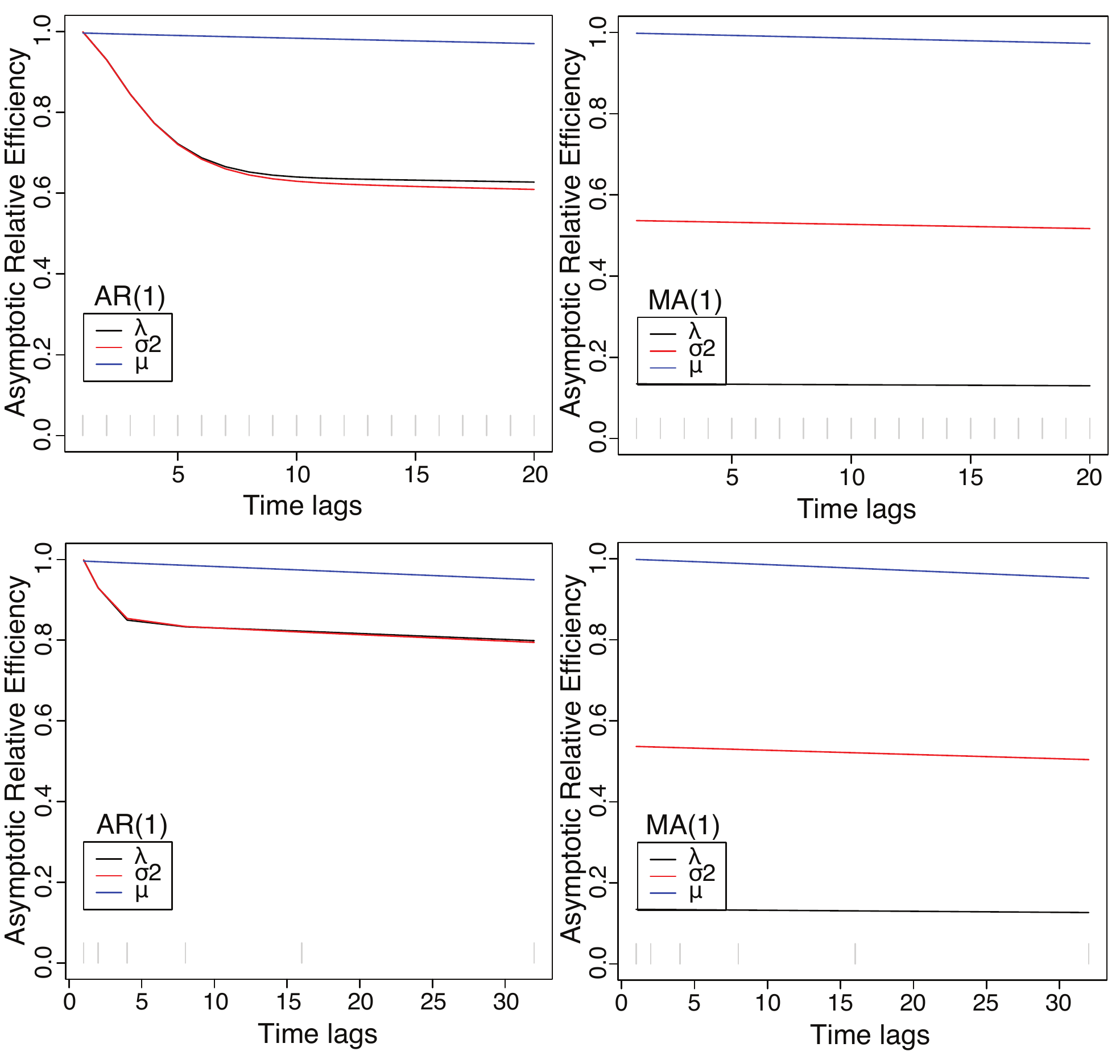}
\caption{Asymptotic efficiency of maximum pairwise likelihood estimators relative to the maximum likelihood estimator, as a function of the maximum time lag included in the pairwise likelihood. The pairwise likelihood of equation (\ref{pairwiselik}) is modified accordingly by setting $S=1$ and replacing $\psi_u$ by the corresponding pairwise density. \emph{Top row}: $\calK_a^K=\{1,\ldots,K\}$. \emph{Bottom row}: $\calK_c^K=\{2^{k-1} ; k=1,\ldots,K\}$. \emph{Left column}: AR$(1)$ process $(Z_t-\mu)=\lambda(Z_{t-1}-\mu) + \v_t$, with $\v_t\iid\calN(0,\sigma^2)$, $\sigma>0,\mu\in\Real,|\lambda|<1$. \emph{Right column}: MA$(1)$ process $Z_t=\mu+\v_t + \lambda\v_{t-1}$, with $\v_t\iid\calN(0,\sigma^2)$, $\sigma>0,\mu\in\Real,|\lambda|<1$. The parameters are $\theta=0.6$, $\sigma^2=1$, and $T=500$.}
\label{Figure1}
\end{figure}

Figure~\ref{Figure1} displays the asymptotic relative efficiency (ARE) of the pairwise likelihood estimator with respect to the maximum likelihood estimator, that is $\avar(\hat\theta_{\rm MLE})/\var(\hat\theta_{p,\calK})$, for different sets $\calK$ of time lags. We considered (a) $\calK_a^K=\{1,\ldots,K\}$, for which all time lags are used up to some maximum time lag $K$; (b) $\calK_b^K=\{b_k : k=1,\ldots,K\}$ where $b_k$ is based on the Fibonacci sequence; and (c) $\calK_c^K=\{2^{k-1} : k=1,\ldots,K\}$ for which the lags increase exponentially. Since the efficiency curves were found to be qualitatively similar for (b) and (c), we only present the results for $\calK_a^K$ and $\calK_c^K$. The left-hand column of Figure~\ref{Figure1} displays the ARE for the AR$(1)$ process, and the results for the MA$(1)$ process are shown in the right-hand column. The top row shows the efficiency curves for $\calK_a^K$ and the bottom row considers the set $\calK_c^K$. As mentioned by \citet{Davis03}, the efficiency is maximized when pairs at lag $1$ only are included. 

In the top left panel (AR$(1)$ and $\calK_a^K$), the ARE for the dependence parameter $\lambda$ is $100\%$ when $\calK=\{1\}$ and then decreases sharply before stabilizing at about lag $9$. This shape is reproduced qualitatively in the bottom left panel, when only the pairs at lags $2^k$ are taken into account, but the efficiency stabilizes at a higher level. However, in practice, one might need to include more distant pairs to ensure parameter identifiability. When the pairs at lags $1,2,3,4,5,6$ are included, the efficiency of the estimator, around $70\%$,  is significantly lower than when the pairs at lags $1,2,4,8,16,32$ are included. Therefore, for a fixed number of pairs, here $6$, it is advantageous to include some distant pairs as well. Thus, for the AR$(1)$, it is better to include not only strongly dependent pairs, but also weakly dependent ones.

The results for the MA$(1)$ process suggest that the efficiency is little affected either by the selection of pairs or by the number of time lags considered, but the ARE is extremely low for the dependence parameter $\lambda$. Other results (not shown) reveal that the efficiency for $\lambda$ drops dramatically as $\lambda$ approaches $\pm 1$, so the loss in ARE is substantial even for moderately correlated MA$(1)$ processes.

When max-stable processes are considered, these results can only be treated as analogies. However, it seems that two main conclusions can be drawn: including many pairs in the pairwise likelihood can spoil the estimator, suggesting that we should retain as few pairs as possible, provided the parameters remain identifiable, and incorporating information from temporally distant (or weakly correlated) pairs is valuable when the process is autoregressive.

\section{Simulation study}\label{SimulationSection}
\begin{figure}[h!]
\centering
\includegraphics[width=16cm]{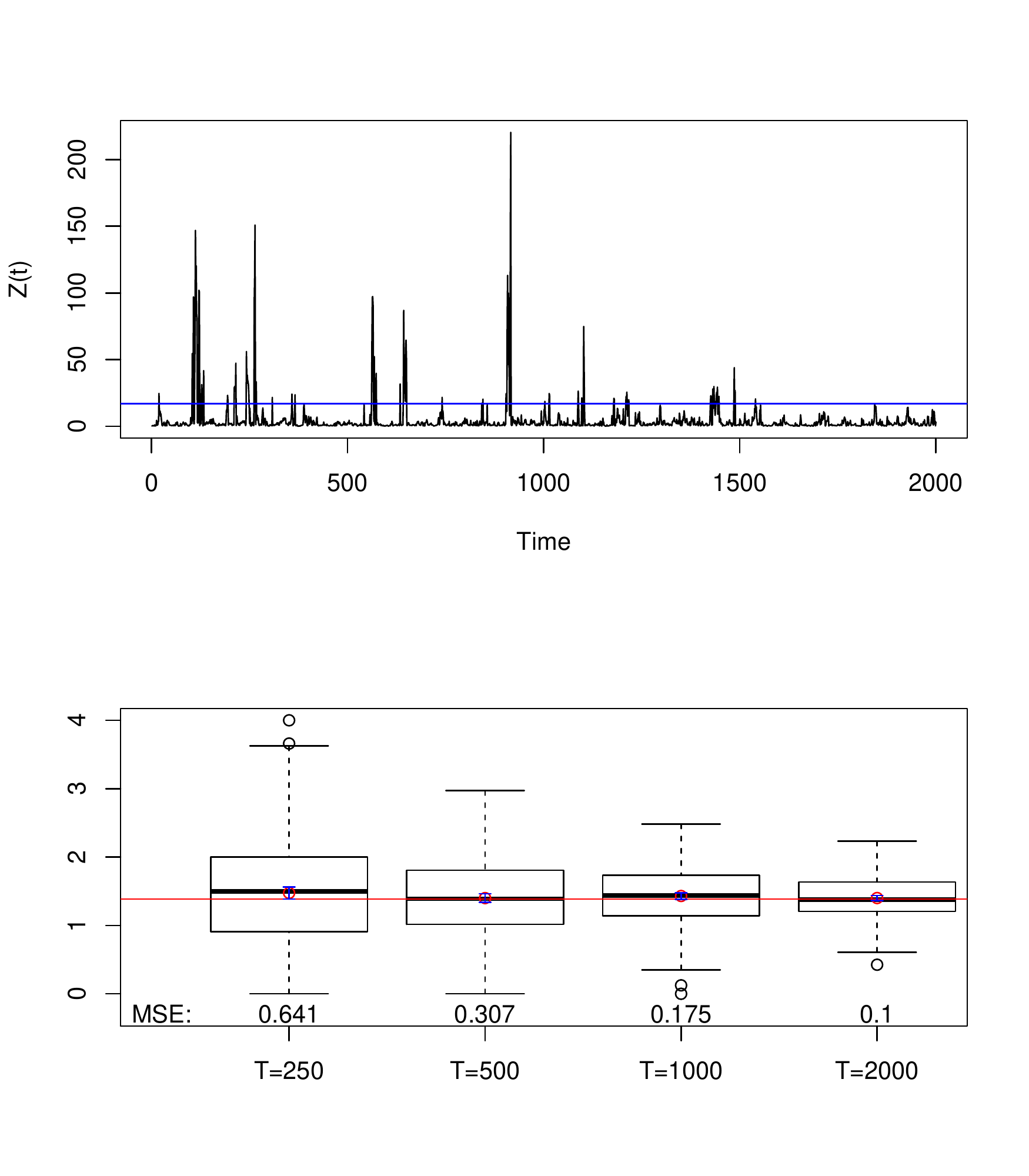}
\caption{\emph{Top}: Simulation of the Schlather model at a particular location with beta distributed random sets. The correlation is exponential with range parameter $\lambda=4$, giving an effective range of $12$. The blue line represents the 0.95-quantile. \emph{Bottom}: Boxplots (with corresponding mean squared errors) of the estimates of $\log\lambda$ (based on $300$ replications) using pairs at lag $1$ only, for an increasing number of observations $T$. 
The true value is the horizontal red line at $\log 4\simeq 1.38$.}\label{Simulation}
\end{figure}

The considerations on efficiency discussed in Section~\ref{EfficiencySection} being based on simple time series models, it is important to check to what extent the conclusions extend to max-stable processes. We therefore conducted a simulation study in a one-dimensional framework. We simulated the Schlather model (\ref{SchlatherModel}) on the time axis, taking $\calX=[0,2000]$, with random sets of the form $\calB=[0,D]$, where $D=24\delta$ and $\delta \sim{\rm beta}(10,240/\mu - 10)$ with $\mu=40/3$, so $\E(D)=\mu\simeq 13.3$. We chose an exponential correlation for the underlying Gaussian random field $\v$, with range parameter $\lambda=4$;  the effective range is $12$. These parameters were chosen to mimic rainfall data. The top panel of Figure~\ref{Simulation} displays a realization from this model.

Fixing the parameter $\mu$ of the random set to its true value, we then estimated the logarithm of the range parameter, $\log\lambda$, with the threshold-based pairwise likelihood estimator of equation (\ref{pairwiselikEst}). We tested different estimators corresponding to the three sets of time lags used in Section~\ref{EfficiencySection}, namely $\calK_{a}^K$, $\calK_b^K$ and $\calK_c^K$, for $K=1,6,9$. Table~\ref{ResultsSim1} reports the mean squared errors (MSE) of these estimates based on $300$ realizations of the Schlather model.

\begin{table}[t!]
\centering
\caption{Mean squared errors (MSE) of the estimates of $\log \lambda$, the logarithm of the correlation range parameter, based on $300$ replications of the Schlather model, for different sets of pairs included in the pairwise likelihood. }\label{ResultsSim1}
\begin{tabular}{|r||c||c|c||c|c|c||c|c|c|}
\hline
 Number of time lags $K$ & $1$ & \multicolumn{2}{|c||}{$3$} & \multicolumn{3}{|c||}{$6$} &  \multicolumn{3}{|c|}{$9$}\\
\hline
Type of time lags set $\calK$ & $\calK_{a/b/c}^K$ & $\calK_{a/b}^K$ & $\calK_{c}^K$ & $\calK_{a}^K$ & $\calK_{b}^K$ & $\calK_{c}^K$ & $\calK_{a}^K$ & $\calK_{b}^K$ & $\calK_{c}^K$ \\
\hline
MSE & $0.100$ & $0.111$ & $0.115$ & $0.145$ & $0.132$ & $0.126$ & $0.180$ & $0.144$ & $0.136$\\
\hline
\end{tabular}
\end{table}

The MSE is minimized for $\calK=\{1\}$, corroborating the findings of Section~\ref{EfficiencySection} for AR$(1)$ or MA$(1)$ processes. Moreover, the MSE is $13\%$ lower when $\calK_c^6$ is used instead of $\calK_a^6$ and $24\%$ lower when $\calK_c^9$ is used instead of $\calK_a^9$, even though the observations separated by more than $24$ time units were independent. Thus the inclusion of some distant, less dependent, pairs can improve inference significantly for fixed $K$.

The bottom panel of Figure~\ref{Simulation} shows that the bias becomes less and less visible and the variance decays more and more as the number of observations $T$ increases, confirming the theoretical 
results established in Section~\ref{InferenceSection}. The simulation suggests that we can estimate the dependence parameter consistently, as expected.

Joint estimation of the correlation range parameter $\lambda>0$ and the mean duration $\mu\in(0,24)$ of the random set is more difficult. Sometimes the estimate of $\mu$ reaches its upper bound; the percentage of successful convergence of the algorithm reported in Table~\ref{ResultsSim2} is only $59\%$ when we choose $\calK=\{1\}$, while it is respectively $83\%$, $96\%$ and $97\%$ for $\calK_a^6$, $\calK_b^6$ and $\calK_c^6$. The estimators including distant pairs in the likelihood outperform those that do not or that only use the most dependent pairs. The same phenomenon is observed when $K=9$, but the difference is less striking than for $K=6$, as expected. In fact, the pairs at lags less than $6$ are probably ineffective to estimate the duration of sets that last $13.3$ time units on average, and that is why $\calK_b^K$ or $\calK_c^K$ are better choices than $\calK_a^K$. The set of time lags $\calK_c^K$ seems slightly better than $\calK_b^K$, in terms of percentage of successful maximizations of the pairwise likelihood. As far as MSE values are concerned, it again seems that the estimators including distant pairs outperform those that use only nearby pairs. Moreover, it seems that the sets of the form $\calK_b^K$ now have slightly smaller MSEs than $\calK_c^K$. To sum up, both estimators that include pairs at lags in $\calK_b^K$ or $\calK_c^K$ behave appreciably better than $\calK_a^K$,  for fixed $K$.

\begin{table}[t!]
\centering
\caption{Mean squared errors (MSE) and percentages of successful maximizations of the pairwise likelihood for the joint estimation of the mean duration $\mu$ of the random set and the logarithm of the range parameter, $\log\lambda$, when different sets of pairs are included in the pairwise likelihood. This simulation is based on $300$ replications of the Schlather model.}\label{ResultsSim2}
\begin{tabular}{|r||c||c|c||c|c|c||c|c|c|}
\hline
 Number of time lags $K$ & $1$ & \multicolumn{2}{|c||}{$3$} & \multicolumn{3}{|c||}{$6$} &  \multicolumn{3}{|c|}{$9$}\\
\hline
Type of time lags set $\calK$ & $\calK_{a/b/c}^K$ & $\calK_{a/b}^K$ & $\calK_{c}^K$ & $\calK_{a}^K$ & $\calK_{b}^K$ & $\calK_{c}^K$ & $\calK_{a}^K$ & $\calK_{b}^K$ & $\calK_{c}^K$ \\
\hline
MSE for $\log\hat\lambda$ & $0.148$ & $0.129$ & $0.113$ & $0.123$ & $0.111$ & $0.105$ & $0.133$ & $0.129$ & $0.119$\\
\hline
MSE for $\hat \mu$ & $34.8$ & $21.5$ & $20.2$ & $14.9$ & $11.3$ & $14.0$ & $12.6$ & $10.1$ & $12.6$\\
\hline
Successful convergence (\%) & $59.3$ & $70.7$ & $76.0$ & $82.7$ & $95.7$ & $97.3$ & $89.3$ & $92.7$ & $94.7$\\
\hline
\end{tabular}
\end{table}

\section{Data analysis}\label{ApplicationSection}
\subsection{Description of the dataset}\label{DataDescription}
\begin{figure}[h!]
\centering
\includegraphics{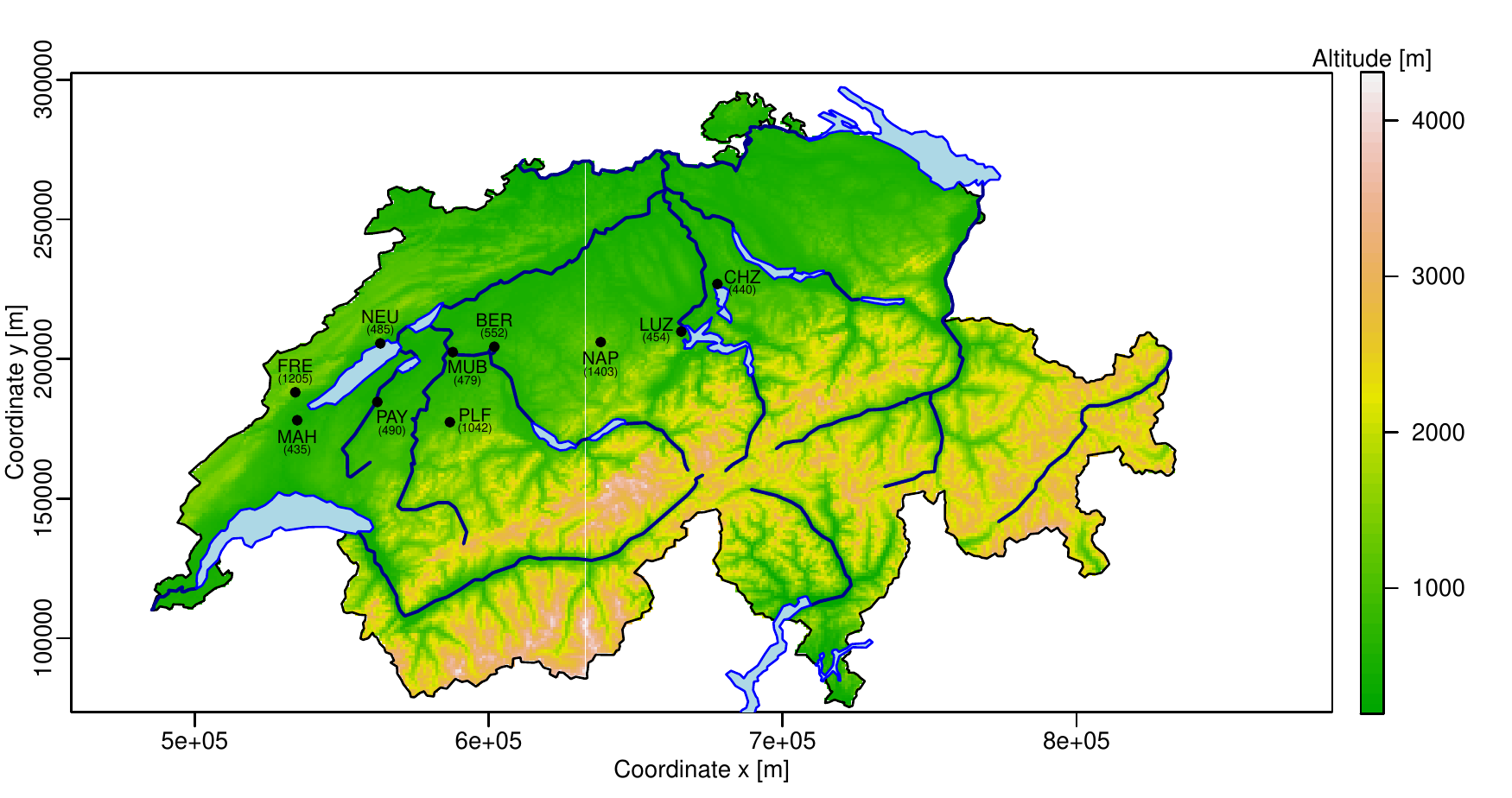}
\caption{Topographic map of Switzerland, showing the location and altitude of the monitoring stations used. Their elevations are all close to $500$ m above mean sea level (amsl), except for three stations (FRE, NAP, PLF) at about $1000$ m amsl. The scales of the $x$ and $y$ axes correspond to the Swiss coordinate system. The closest stations (FRE, MAH) are $10$ km apart and the most distant ones (CHZ, MAH) are $151$ km apart.}\label{Switzerland}
\end{figure}

\begin{landscape}
\begin{figure}[h!]
\centering
\includegraphics[width=23.5cm]{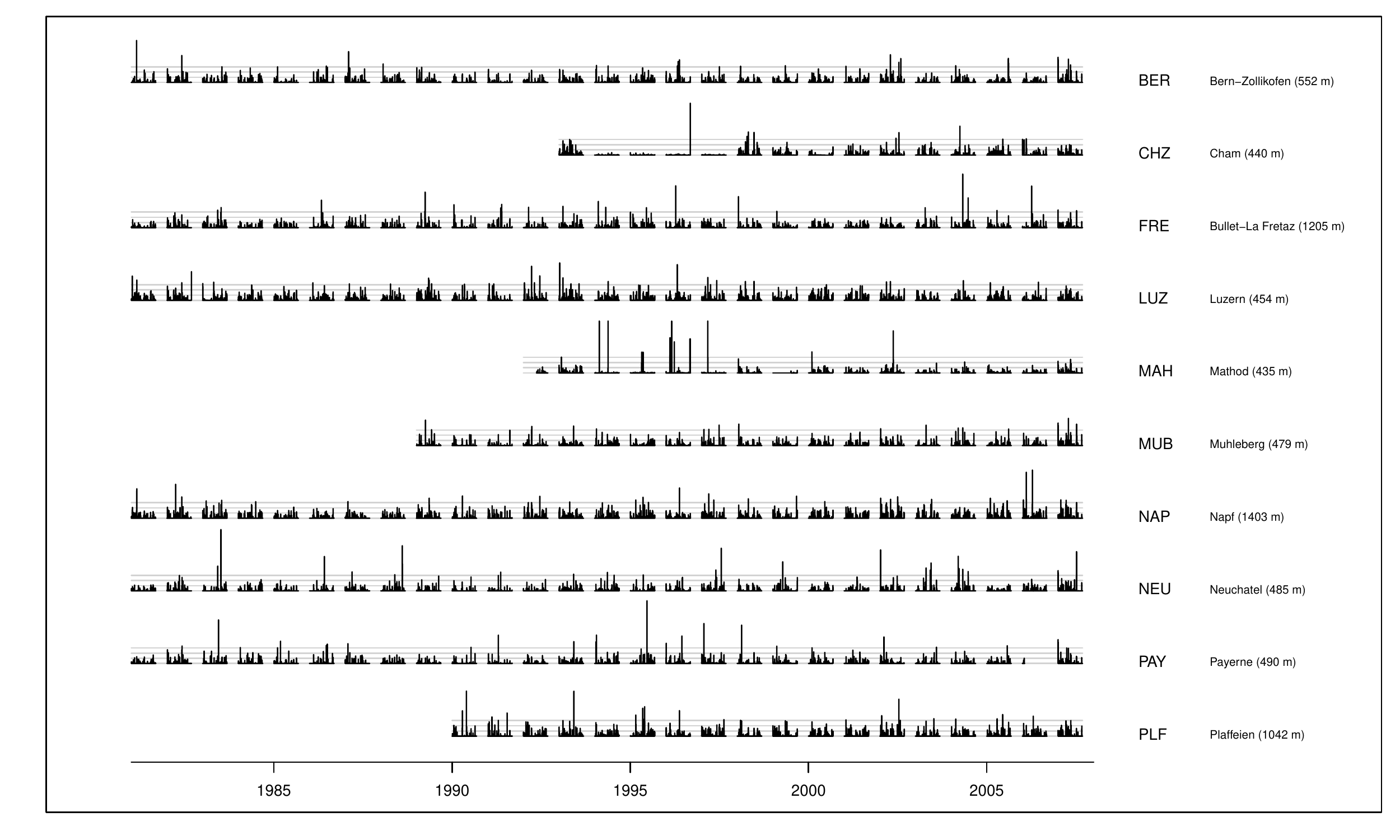}
\caption{Summer hourly rainfall data (mm) at $10$ monitoring stations. The light grey lines show $0, 5, 10,\ldots $ mm. $75\%$ of the measurements equal zero. The univariate thresholds used for transformation to the unit Fr\'echet scale are the 0.97-quantiles, ranging from $0.7$ mm to $1.9$ mm depending on the station. The gaps indicate that summers were treated as independent from one year to the next.}\label{Data}
\end{figure}
\end{landscape}

The dataset used for our application is composed of hourly rainfall measurements (mm) recorded from $1981$ to $2007$ at ten monitoring stations in western Switzerland. Figure~\ref{Switzerland} illustrates the location and topography of the area of study. All stations are located between the Alps and the Jura mountains, and their altitude is almost constant. Only the periods from midnight on June 21st to 11 pm on September 20th were considered, summers being treated as mutually independent. The entire dataset comprises $503988$ measurements, with up to  $59616$ data points per site. The rainfall time series, shown in Figure~\ref{Data}, were independently transformed to the unit Fr\'echet scale, following Section~\ref{UnivariateEVT}, with quantile-quantile plots showing satisfactory agreement between the empirical and fitted quantiles. The thresholds were the 0.97-quantiles of each series. Due to the size of the dataset at each site, the margins were fitted with negligible variability. Below we focus on the modelling of extremal dependence, rather than on the marginal behaviour.

\begin{landscape}
\begin{figure}[h!]
\centering
\includegraphics[height=6in]{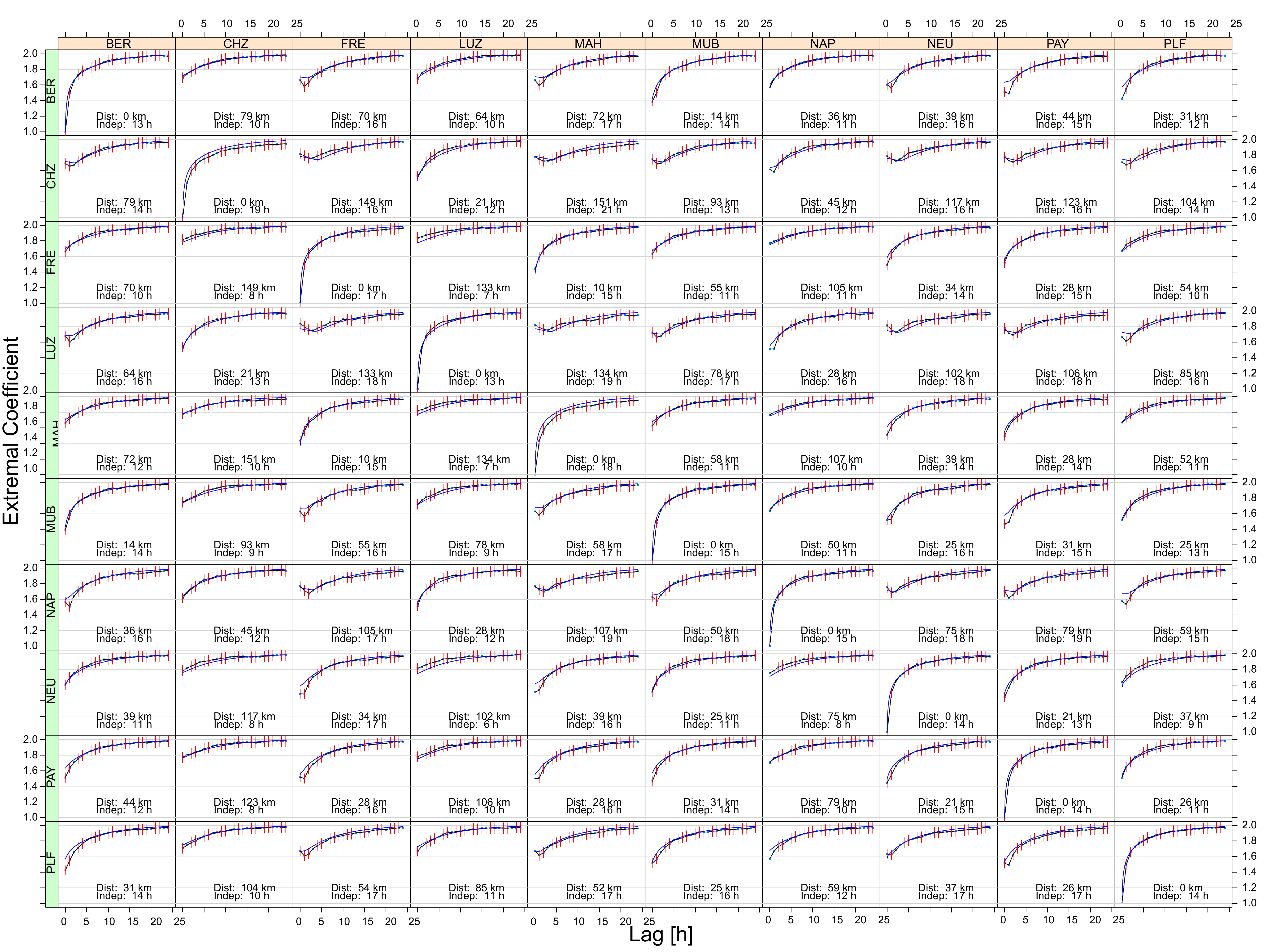}
\caption{Empirical and theoretical pairwise extremal coefficients $\theta_2$ for all pairs of stations. The black lines join the empirical extremal coefficients found using the censored Schlather--Tawn estimator at the 0.95-quantile threshold, the vertical red segments being $95\%$ confidence intervals. The blue lines correspond to the extremal coefficient curves derived from the fitted model. The panel at the \rth row and \cth column shows the extremal coefficients between $Z^{\rm  c}_t$ and $Z^{\rm r}_{t+h}$, for $h=0,1,2,\ldots,24$. ``Dist'' stands for the distance between stations, and ``Indep'' is the time needed to get independence (the first lag for which the value $\theta_2=2$ lies within the confidence interval).}\label{ExtrCoeff}
\end{figure}
\end{landscape}

Figure~\ref{ExtrCoeff} gives an overview of the empirical pairwise extremal coefficients for all pairs of stations at different time lags, based on a censored version of the naive Schlather--Tawn (2003) estimator. There is evidence of significant spatial and temporal dependence between the different series. Panel $(1,1)$ shows the temporal extremal coefficients at Bern-Zollikofen; it starts with the value $1$ (complete dependence at lag $0$), and seems to tend smoothly to the value $2$ (independence) as the time lag increases. This pattern repeats itself for the other sites. The off-diagonal panels represent extremal coefficients for the different pairs of stations, and hence display space-time interactions. For example, Panel $(1,4)$, in the $1{\rm st}$ row and $4{\rm th}$ column, displays the extremal coefficients between the rainfall time series at Luzern at time $t$ and the rainfall time series at Bern-Zollikofen at time $t+h$, for $h=0,1,\ldots$,24. Panel $(4,1)$ reverses the role of the stations. The extremal coefficient functions differ for the panels, showing that the orientation of the stations matters. The extremal coefficient dips towards the value $1$ at lags $1$ or $2$ when the stations are west-east oriented: during the summer months, western Switzerland is governed by dominant winds from the west or north-west, so that the clouds tend to discharge their rain first on the western part of Switzerland.  The same rainfall event could therefore be recorded by two distant monitoring stations at a lag of $1$ or $2$ hours, depending on their location and on the wind velocity. Consequently, extremal dependence might be higher at lag $1$ or $2$ than at lag $0$. A model for the data should be able to capture such features.

\subsection{Model construction}
We now discuss the construction of a model based on (\ref{SchlatherModel}) for the rainfall data described in Section~\ref{DataDescription}. This space-time model comprises a standard Gaussian random field $\v(x)$ with correlation function $\rho(h)$ and a random set element $\calB$, both defined on a space $\calX=\calS\times\calT$, where $\calS=\Real^2$ denotes space and $\calT=\Real_+$ denotes time.

The Gaussian random field is supposed to model the short-range behaviour of the process within single storms, so it is important to have a correlation function that can flexibly capture space-time interactions. For a good review of space-time correlation functions and a discussion of properties such as stationarity, separability and full symmetry, see \citet{Gneiting02} and the references therein.  \citet{Cressie02} propose classes of nonseparable spatio-temporal stationary covariance functions based on Bochner's theorem, and \citet{Gneiting01} extends their work by providing other very general flexible space-time covariance models. \citet{Davis04} show that this class of covariance functions satisfies a natural smoothness property at the origin, directly linked to the smoothness of the random field, and is therefore suitable for the modelling of physical processes such as rainfall. We used the isotropic nonseparable space-time correlation function \citep{Gneiting01}
\begin{equation}\label{CorrelationFunction}
\rho(s,t)={1\over\left\{\left(t\over \exp\alpha_t\right)^{\beta_t} +1\right\}^{d\gamma/2}}\exp\left[-{\left(s\over\exp\alpha_s\right)^{\beta_s}\over\left\{\left(t\over \exp\alpha_t\right)^{\beta_t} +1\right\}^{\beta_s\gamma/2}}\right],
\end{equation}
where $s$ and $t$ are respectively distances in space and time, $\alpha_s,\alpha_t\in\Real$ determine spatial and temporal scale parameters, $\beta_s,\beta_t\in(0,2)$ are spatial and temporal shape parameters, $d=2$ is the spatial dimension, and $\gamma\in(0,1)$ is a separability parameter quantifying the space-time interactions. As $\gamma$ approaches $1$, the spatial and temporal components are less and less separable.

\begin{figure}[h!]
\centering
\includegraphics[width=7.5cm]{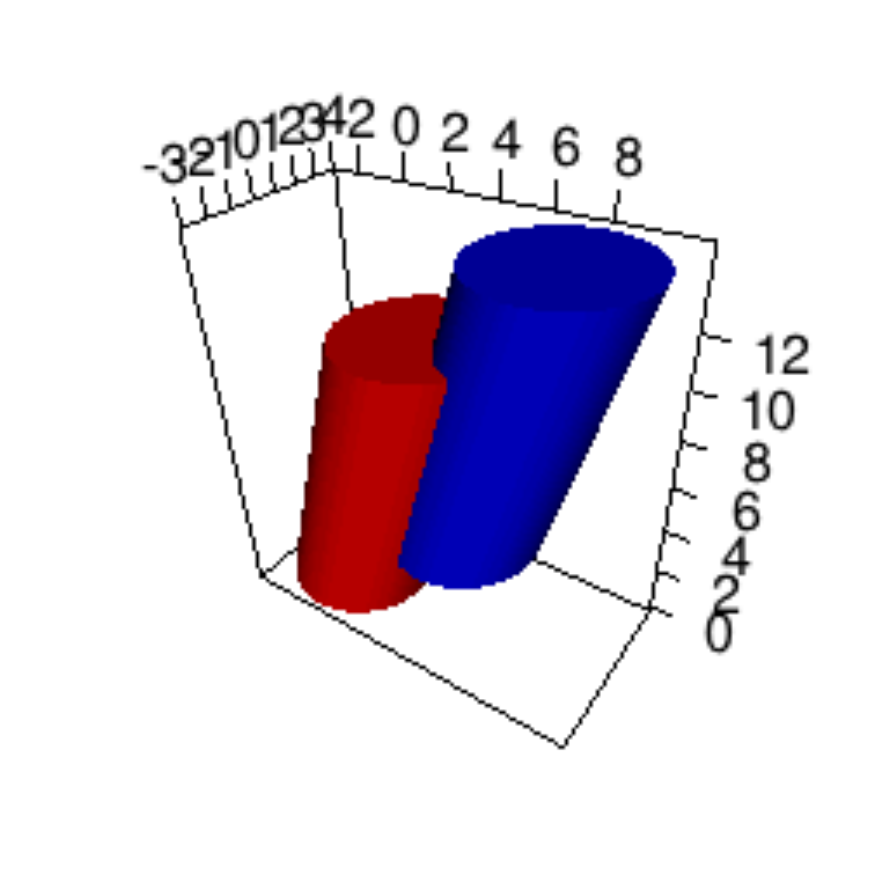}
\caption{Illustration of the random set element $\calB$ in space $\calS$ (horizontal plane) and time $\calT$ (vertical axis). The storms are conceptualized as random disks with a random radius moving at a random velocity for a random duration. The red tilted cylinder represents a realization $\calB$ of such a storm in $\calS\times\calT$, and the blue one is $\calB+h$, for a given vector $h$. The coefficient $\alpha(h)$ needed for the fitting is the expected volume of intersection between the two cylinders.}
\label{RandomSet}
\end{figure}

The random set $\calB$ is interpreted as a random storm having a finite extent, which enables the model to capture complete independence. Conceptualizing storms as disks of random radius $R$ moving at a random velocity $V$ for a random duration $D$ starting from a random position, the storm extent $\calB$ in space and time becomes a tilted cylinder in $\calS\times\calT$, with a Schlather process inside; see Figure~\ref{RandomSet}. For tractability we assume that $R\sim{\rm Gamma}(m_R/k_R,k_R)$ (with mean $m_R$ km),  $V\sim\calN_2(m_V,\Omega)$ (km/hour) and  $D\sim{\rm Gamma}(m_D/k_D,k_D)$ (with mean $m_D$ hours).

\subsection{Model fitting}
The fitting of our model requires the computation of the coefficient $\alpha(h)=\E\{|\calB\cap(h+\calB)|\}/\E(|\calB|)$ for $h\in\calX$, i.e.,  the normalized expected volume of overlap between the random set $\calB$ and itself shifted by the space-time lag $h$. Several mild approximations, some analytical calculations and a single one-dimensional finite integration yield a good approximation to $\alpha(h)$; see  Appendix~\ref{ComputationOfAlpha}.

After some exploratory analysis we fixed $\beta_t=1$, and then the model has four parameters for the correlation function, and nine for the random set. Due to the complexity of the problem, we split the estimation procedure into four parts: we estimate first the temporal parameters ($\alpha_t,m_D,k_D$), then the spatial parameters ($\alpha_s,\beta_s,m_R,k_R$), then the spatio-temporal parameters ($\gamma,m_V,\Omega$) with the other parameters held fixed to their estimates, and finally all the parameters together, with the former estimates as starting values. We always use the pairwise likelihood estimator (\ref{pairwiselikEst}). Standard errors are calculated by the block jackknife (see Section~\ref{VarianceEstimation}), using yearly blocks. Based on the results in Sections~\ref{EfficiencySection} and \ref{SimulationSection}, we include  the pairs at lags in $\calK=\{0,1,2,4,8,16\}$ in the pairwise likelihood. A single evaluation involves contributions for about $T|\calK||S|^2=50000\times 6\times 10^2 = 30$ million pairs, while the full pairwise likelihood would have $7$ billion pairs, completely impractical for inference purposes! We coded the pairwise likelihood in C, parallelized the work on $8$ CPUs
, and fitted the model using the R optimization routine L-BFGS-B with specific box constraints. Due to the complex model and the amount of data, a single full estimation took 5 days. As the $27$ bootstrap replicates can be computed independently, $27\times8 = 216$ CPUs were used simultaneously to estimate the standard errors. The results are presented in Table~\ref{Results}. 

\begin{table}[t!]
\centering
\caption{Parameter estimates and standard errors from fitting our random set model to the rainfall data.}\label{Results}
\begin{tabular}{llllr}
  \hline
 & & & & Estimate (SE)\\
   \hline
Correlation &Scale & Space & $\exp(\alpha_s)$ (km)& 35.5~(5.97)\\
& & Time &$\exp(\alpha_t)$ (hr)& 1.00~(0.14)\\
&Shape& Space &$\beta_s$ & 0.98~(0.06)\\
& & Time &$\beta_t$ & 1~(---)\\
&Separability& &$\gamma$ & $0.99$~(0.00)\\
\hline
  Random Set &Duration&Mean&$m_D$ (hr)& 36.78~(0.34)\\
& &Shape&$k_D$& 9.75~(0.01)\\
&Radius&Mean&$m_R$ (km)& 51.21~(0.16)\\
& &Shape&$k_R$& 0.28~(0.05)\\
&Velocity&Mean&$m_V$ (km/hr)& 32.67~(0.74)\\
& & & & 11.41~(0.16)\\
& &Standard Deviations&$\Omega^{1/2}_{11}$ (km/hr)& 3.00~(0.01)\\
& & &$\Omega^{1/2}_{22}$  (km/hr)& 3.43~(0.00)\\
& &Correlation&$\rho_{12}$& $-0.95$~(0.03)\\
   \hline
\end{tabular}
\end{table}

The estimated mean duration $m_D$ and mean radius $m_R$ of a storm are $37$ hours and $51$ km, which seem reasonable when compared to radar images of precipitation for the same region and time of year. The mean velocity $m_V$ has components $33$ and $11$ km/hr, so the mean speed is about $34.6$ km/hr and the angle of the dominant winds is about $19^\circ$ in the Argand diagram. This means that the clouds are likely to move from west to east (and slightly to the north), in agreement with the dynamic of the oceanic climate in Western Switzerland during the summer. The correlation $\rho_{12}$ of the velocity is close to $-1$, so the angle of the velocity is less easily determined than its length.

The separability parameter $\gamma$ always reached the bound $0.99$, suggesting that the data are highly nonseparable and that this parameter tries to capture it.

Overall the standard errors seem very small despite the large amount of data available. The use of monthly (instead of yearly) blocks produced similar standard errors, but this could be due to the inappropriateness of the bootstrap procedure or the instability of the pairwise likelihood around the maximum pairwise likelihood estimate. For computational reasons we could not investigate this further, so the standard errors should be interpreted with care.

\subsection{Model checking}
\begin{figure}[h!]
\centering
\includegraphics[width=18cm]{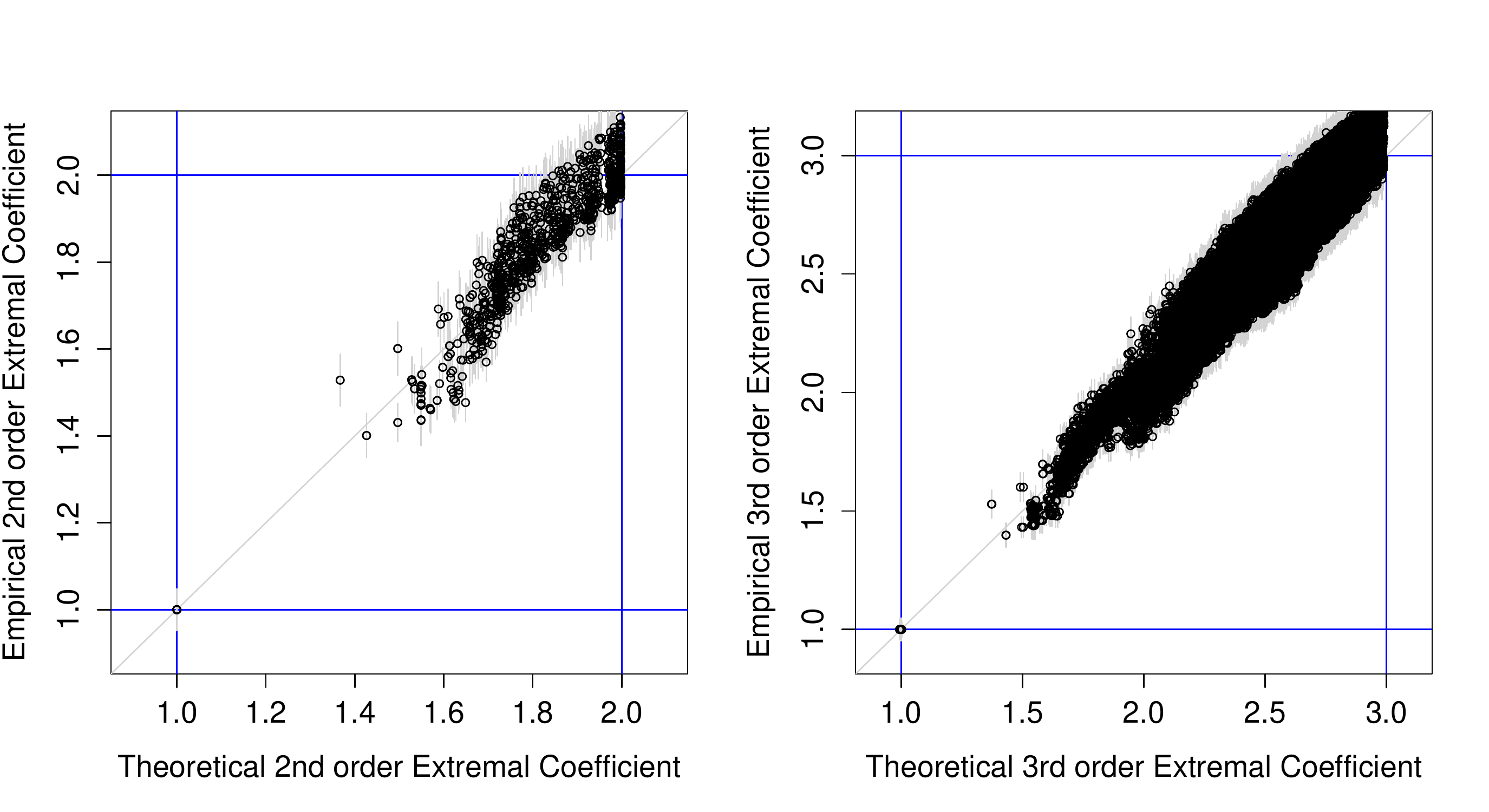}
\caption{Comparison of empirical estimates of all pairwise (left) and trivariate (right) extremal coefficients for the rainfall data with their model-based counterparts. The light-grey vertical lines are $95\%$ confidence intervals. A perfect agreement would place all points on the grey diagonal line.}\label{PairTriExtrCoeff}
\end{figure}

Figure~\ref{ExtrCoeff} compares empirical estimates of the pairwise extremal coefficients with their model-based counterparts. There is a good agreement overall, but the model systematically underestimates extremal dependence at lag $1$. This lack of fit at short time lags can be explained either by a lack of flexibility due to the (conceptually) simplistic model that we used or by the difficulty to reach the global pairwise likelihood maximum for such a model. The diagonal plots, showing the marginal dependence of the extremes, show a good fit. The small differences at Cham (CHZ) or Mathod (MAH) may be due to nonstationarity or because data at those monitoring stations might not have been cleaned properly; see Figure~\ref{Data}. The left panel of Figure~\ref{PairTriExtrCoeff} shows the pairwise extremal coefficients $\theta_2$ in (\ref{ExtrCoeffDefinition}).

As the model was fitted with pairs of observations, one might wonder whether it can capture higher-order interactions. We therefore computed the trivariate extremal coefficients (see Appendix~\ref{TriExtrCoeffDetails}) and found good agreement between nonparametric estimates of trivariate extremal coefficients and their model-based counterparts; see Figure~\ref{PairTriExtrCoeff}. Despite the strong dependence among these trivariate estimates and with the pairwise counterparts, it seems that the trivariate interactions are fairly well modelled. The biggest discrepancies are from stations CHZ (Cham) and MAH (Mathod), but without these stations, the points are well concentrated around the diagonal.

In order to assess the sensitivity of the results to initial conditions, we re-fitted the model with different starting values. The results were sometimes fairly different, but with similar bivariate properties and with almost the same likelihood. Consequently, we believe that some parameters are likely to play a similar role, giving rise to identifiability issues. Indeed, our stations are at most $150$ km apart, which is not very distant, if a cloud of radius $50$ km moves at about $35$ km/hr.


\section{Discussion}\label{DiscussionSection}

The work described above extends existing statistics of extremes by proposing a flexible class of models for spatio-temporal extremes, applied here to rainfall, but with clear possibilities for extension to other phenomena. `Dynamic' space-time modelling of extremes thus seems to be feasible;  complex models can be consistently fitted using composite censored likelihood based on threshold exceedances.  However, the large amount of data and the consequent use of parallel computation underlines the necessity for substantial computing resources when tackling such problems: in our application, the fitting would otherwise have been completely out of reach. 
%

Although highly idealized, our model is still fairly complex, and estimation and simulation are demanding. Moreover, the assessment of fit is tricky, due to the computational burden that it requires. After a major effort we were able to check the trivariate interactions by means of the {$3{\rm rd}$ order} extremal coefficients, and although it would be feasible to use simulation to investigate higher order interactions, it would be awkward.

An important modelling issue is that near-independence cannot be captured by our model, which is based solely on max-stable processes. However, it is common in practice to observe two distinct events becoming less and less dependent as their rarity increases. \citet{Wadsworth01} have proposed models that can handle both asymptotic independence and asymptotic dependence, and it seems entirely feasible to extend our approach to them.


\section*{Acknowledgement}\label{AcknowledgementSection}
This research was funded by the Swiss National Science Foundation, and partly performed in the context of the Competence Center Environment and Sustainability (CCES).

\bibliographystyle{CUP.bst}
\bibliography{references.bib}

\newpage

\section*{Appendix}
\begin{appendix}

\section{Proof of Theorem 1}\label{ProofOfThm1}
\begin{small}
\begin{proof}
For notational simplicity, we give the proof in the case where the parameter $\theta$ is scalar, but the argument can be extended to the vector case.

By definition of the pairwise likelihood in equation~(\ref{pairwiselik}), and as the observations $z_{s,t}$ are realizations of a max-stable process, we have
$$\E\{U_t(\theta)\}=\sum_{h\in\calK_t}\sum_{s_2=1}^S\sum_{s_1=1}^S \underbrace{\E\left\{{d\over d\theta}\log p_u\left(z_{s_1,t},z_{s_2,t+h};\theta\right)\right\}}_{=0}(1-I\{s_1\geq s_2 \mbox{ and } h=0\}).$$
Therefore, we also have that $\E\{U(\theta)\}=\E\{\sum_{t=1}^T U_t(\theta)\}=0$.

In addition, the variance of $U(\theta)$ renormalised by $T$ is \citep[p.510]{Shumway01}
\begin{eqnarray*}
T^{-1}\var\{U(\theta)\}&=&\E\{U_1(\theta)^2\} + 2\sum_{t=1}^{T-1}\left(1-{t\over T}\right)\E\{U_1(\theta)U_{t+1}(\theta)\}\\
&\to &\E\{U_1(\theta)^2\} + 2\sum_{t=1}^{\infty}\E\{U_1(\theta)U_{t+1}(\theta)\},\quad\mbox{as $T\to\infty$, if the sum converges absolutely.}
\end{eqnarray*}
Now, as $\hat\theta_{p,\calK}$ is the maximum pairwise likelihood estimator, second-order Taylor expansion of $U_t(\hat\theta_{p,\calK})$ around the true parameter $\theta$ gives 
$$0=\sum_{t=1}^T U_t(\hat\theta_{p,\calK}) \doteq \sum_{t=1}^T \left\{U_t(\theta) + {d\over d\theta} U_t(\theta)(\hat\theta_{p,\calK}-\theta)\right\},$$
which gives, up to a term of the order $O\{(\hat\theta_{p,\calK}-\theta)^2\}$, that 
\begin{equation}\label{TaylorExp}
\hat\theta_{p,\calK}\doteq \theta + \left\{\sum_{t=1}^T H_t(\theta)\right\}^{-1}\sum_{t=1}^T U_t(\theta) = \theta + H(\theta)^{-1}U(\theta),
\end{equation}
where $H_t(\theta)=-d U_t(\theta)/d\theta$ and $H(\theta)=\sum_{t=1}^T H_t(\theta)$ is the observed information.
Moreover, since the process $Z(x)$ is assumed to be temporally $\alpha$-mixing with coefficients $\alpha(n)$, the time series $U_t(\theta)$ is also $\alpha$-mixing with coefficients $\alpha'(n)=\alpha(n-\max\calK)$. Hence, 
$$\alpha'(n)\to 0,\qquad\qquad \sum_{n\geq 1}|\alpha'(n)|^{\delta/(2+\delta)}<\infty,$$
with the same $\delta>0$. These results, along with the assumptions $\E(U_1^2)<\infty$ and $\E(|U_1|^{2+\delta})<\infty$, ensure that the Central Limit Theorem $10.7$ of \citet{Bradley01} applies, and thus
$$
T^{-1/2}U(\theta)\Dto \calN\{0,K(\theta)\},\quad T\to\infty, 
$$
where $K(\theta)=\E\{U_1(\theta)^2\} + 2\sum_{t=1}^{\infty}\E\{U_1(\theta)U_{t+1}(\theta)\}<\infty$ and $\Dto$ denotes convergence in distribution. Therefore, coming back to equation~(\ref{TaylorExp}), and by definition of $J_1(\theta)$, by the law of large numbers, and by Slutsky's theorem, we get
\begin{eqnarray*}
T^{1/2}(\hat\theta_{p,\calK}	-\theta) & \doteq & T^{1/2}H(\theta)^{-1} U(\theta) \\
& = & \{T^{-1}H(\theta)\}^{-1} \{T^{-1/2}U(\theta)\}\\
& \Dto & J_1(\theta)^{-1}\calN\{0,K(\theta)\}\qquad\mbox{as $T\to\infty$}\\
& \Deq & \calN(0,J_1(\theta)^{-1}K(\theta)J_1(\theta)^{-1}),
\end{eqnarray*}
where $\Deq$ denotes equality in distribution. 
But $K(\theta)$ is the asymptotic variance of the score, renormalized by $T$. Hence, the result is proved.
\end{proof}
\end{small}

\section{Computation of the volume of overlap $\alpha(h)$}\label{ComputationOfAlpha}
\begin{small}
The coefficient $\alpha(h)$ is defined as $\E\{|\calB\cap(h+\calB)|\}/\E(|\calB|)$, where $\calB$ is a tilted cylinder in $\calX=\calS\times\calT=\Real^2\times\Real_+$ (see Figure~\ref{RandomSet}), and $h=(s,t)\in\calX$. If the cylinder were vertical (zero wind velocity), the volume of overlap would simply be the product of the area of overlap between two discs distant by $\|s\|$ and the corresponding height, the storm duration minus $t$. 

Let $R$ be the storm radius, $V=(V_1,V_2)\in\Real^2$ be its velocity and $D$ be its lifetime. A good linear approximation to the area of overlap of two discs of radius $R$ distant by $d$ is $\pi R^2\max\{0,1-d/(2R)\}$ \citep{Davison03}. Therefore, for a vertical cylinder $\calB$, $|\calB\cap(h+\calB)|$ can be approximated by
$$\pi R^2\left(1-{\|s\|\over 2R}\right)_+ (D-t)_+,$$
where $a_+=\max\{0,a\}$. When the cloud is moving, giving a tilted cylinder, a simple geometric argument can be used to prove that in the general case, the volume of overlap is transformed to
$$|\calB\cap(h+\calB)|\doteq\pi R^2\left(1-{d^*\over 2R}\right)_+ (D-t)_+,$$
where $d^*=\left[\|s\|^2 + t^2(V_1^2 + V_2^2) - 2\|s\|t\{V_1\cos(\theta)+V_2\sin(\theta)\}\right]^{1/2}$, $\theta=\arctan(s_1/s_2)$ being the angle between the stations with respect to a reference axis in the West-East direction. In order to compute the coefficient $\alpha(h)$, which depends upon the spatial distance $\|s\|$, the temporal lag $t$ and the orientation of the stations $\theta$, we need to obtain the expected volume of overlap $\E\{|\calB\cap(h+\calB)|\}$, by putting tractable distributions on $R$, $D$, and $V=(V_1,V_2)$. We choose to set
\bi
\item $R\sim{\rm Gamma}(m_R/k_R,k_R)$ (with mean $m_R$ km);
\item $V\sim\calN_2(m_V,\Omega)$ (km/hour), with $m_V=(m_1,m_2)^T$ and $\Omega=\begin{pmatrix}
\sigma_1^2 & \sigma_1\sigma_2\rho_{12}\\
\sigma_1\sigma_2\rho_{12} & \sigma_2^2
\end{pmatrix}$;
\item $D\sim{\rm Gamma}(m_D/k_D,k_D)$ (with mean $m_D$ hours),
\ei

\noindent and we assume that $R$, $D$ and $V$ are mutually independent. To compute this expectation, note first that $(D-t)_+$ can be integrated out analytically. Second, by conditioning on $V$, it is possible to integrate over $R$ as well. We can then reduce the full computation to this single expectation with respect to $V=(V_1,V_2)$:
\begin{equation}\label{alphaH}
\alpha(h) = \E_V\left\{\pr(G_{m_R/k_R;k_R+2} > d^*/2) - {d^* k_R\over 2 (k_R+1)m_R}\pr(G_{m_R/k_R;k_R+1} > d^*/2)\right\},
\end{equation}
where $G_{\theta;k}$ is a gamma random variable with scale parameter $\theta$ and shape parameter $k$; its mean equals $m=\theta k$. Expression (\ref{alphaH}) does not have a closed form, but it can be remarkably well approximated by a function of the form
$$e^{-a\sqrt{(V_1-\mu_1)^2 + (V_2-\mu_2)^2}},$$
where $a$ is real number which does not depend upon $V=(V_1,V_2)$, and that can be estimated with a few points by least squares, and where $\mu_1=\|s\| \cos(\theta) / t$ and $\mu_2=\|s\| \sin(\theta) / t$. Therefore, we have
\begin{eqnarray}
\alpha(h)&\approx&\E_V\left\{e^{-a\sqrt{(V_1-\mu_1)^2 + (V_2-\mu_2)^2}}\right\}\nonumber\\
&=&\int_{\Real^2} e^{-a\sqrt{(v_1-\mu_1)^2 + (v_2-\mu_2)^2}} {1\over 2\pi{\rm det}(\Omega)^{1/2}} e^{-{1\over 2}(v_1 - m_1;v_2-m_2)\Omega^{-1}(v_1 - m_1;v_2-m_2)^T} dv_1dv_2\nonumber\\
&=&{1\over 2\pi {\rm det}(\Omega)^{1/2}}\int_{\Real^2} e^{-a\sqrt{(v_1-\mu_1)^2 + (v_2-\mu_2)^2} - {1\over 2{\rm det}(\Omega)}\left\{(v_1-m_1)^2\sigma_2^2 - 2(v_1-m_1)(v_2-m_2)\sigma_1\sigma_2\rho_{12} +(v_2-m_2)^2\sigma_1^2\right\}}dv_1dv_2\nonumber\\
&=&{1\over 2\pi {\rm det}(\Omega)^{1/2}}\int_0^{2\pi}d\xi\int_{\Real_+} re^{-ar - {1\over 2{\rm det}(\Omega)}\{r^2 a(\xi) + rb(\xi) + c(\xi)\}}dr\label{expr1}\\
&=&{1\over (2\pi)^{1/2}}\int_0^{2\pi}{1\over \sqrt{a(\xi)}} e^{-{1\over 2\sigma(\xi)^2} \left({{c(\xi)\over a(\xi)} - \mu(\xi) ^2 }\right)}d\xi\int_{\Real_+} r {1\over \sqrt{2\pi} \sigma(\xi)} e^{-{1\over 2}\left({r-\mu(\xi)\over \sigma(\xi)}\right)^2} dr\nonumber\\
&=& {1\over 2\pi}\int_0^{2\pi} {1\over \sqrt{a(\xi)}} e^{-{1\over 2\sigma(\xi)^2} \left({{c(\xi)\over a(\xi)} - \mu(\xi) ^2 }\right)}\left[\sigma(\xi) e^{-{1\over 2}{\mu(\xi)^2\over \sigma(\xi)^2}} + \sqrt{2\pi}\mu(\xi)\left\{1-\Phi\left(-{\mu(\xi)\over\sigma(\xi)}\right)\right\}\right] d\xi,\label{expr2}
\end{eqnarray}
where $\Phi(\cdot)$ is the normal cumulative distribution function and
\begin{eqnarray*}
a(\xi) &=& \cos^2(\xi)\sigma_2^2 + \sin^2(\xi)\sigma_1^2 - 2\cos(\xi)\sin(\xi)\sigma_1\sigma_2\rho_{12}\\
b(\xi) &=& 2\cos(\xi)(\mu_1-m_1)\sigma_2^2 + 2\sin(\xi)(\mu_2-m_2)\sigma_1^2-2\cos(\xi)(\mu_2-m_2)\sigma_1\sigma_2\rho_{12} - 2\sin(\xi)(\mu_1-m_1)\sigma_1\sigma_2\rho_{12}\\
c(\xi)&=&(\mu_1-m_1)^2\sigma_2^2 + (\mu_2-m_2)^2\sigma_1^2 - 2(\mu_1-m_1)(\mu_2-m_2)\sigma_1\sigma_2\rho_{12}\\
\mu(\xi)&=&-{b(\xi)\over 2 a(\xi)}- {a{\rm det}(\Omega) \over a(\xi)},\qquad\qquad\sigma(\xi)\;\;=\;\; \sqrt{{\rm det}(\Omega) / |a(\xi)|},\qquad\qquad{\rm det}(\Omega) \;\;=\;\; \sigma_1^2\sigma_2^2 (1- \rho_{12}).
\end{eqnarray*}
Expression (\ref{expr1}) above was computed with a straightforward change of variables $v_1= r\cos(\xi) + \mu_1$, $v_2=r\sin(\xi) + \mu_2$, while expression (\ref{expr2}) stems from the properties of the normal cumulative function. Since the integral (\ref{expr2}) is impossible to handle analytically, we can use a finite approximation to estimate $\alpha(h)$, based on $100$ points equi-spaced in the interval $[0,2\pi]$. The approximation seems to be  adequate when $\sigma_1^2,\sigma_2^2>5$, which we impose in the R optimization routine L-BFGS-B.

\end{small}

\section{Trivariate extremal coefficients for model (\ref{SchlatherModel})}\label{TriExtrCoeffDetails}
\begin{small}
From equation~(\ref{ModelForMaxima}), we know that the multivariate extremal coefficient in dimension $N$ is
$$
\theta_N=V_N(1,\ldots,1)=\E\left[\max_{i=1,\ldots,N}\{W(x_i)\}\right].$$
This takes values between $1$ and $N$, ranging from complete dependence to asymptotic independence. Therefore, the extremal coefficient of order $3$ is 
$$\theta_{3}=\E\left[\max\{W(x_1),W(x_2),W(x_3)\}\right],$$
where, for model (\ref{SchlatherModel}), $W(x)\propto\max\{0,\v(x)\}I_{\calB}(x-X)$, $x\in\calX$, $\v(x)$ being an isotropic Gaussian random field with zero mean, variance $1$ and correlation function $\rho(h)$ and $I_\calB$ being the indicator that the point $x-X$ belongs to a random set $\calB$ (where $X$ is a Poisson process in $\calX$). The proportionality constant is such that $W(x)$ has mean $1$, so it must be 
$${1\over\E\{\max(0,\v)I_\calB\}}={1\over\E\{\max(0,\v)\}\E(I_\calB)}={\sqrt{2\pi}\over\pr(x\in\calB)}={\sqrt{2\pi} |\calX|\over\E(|\calB|)}.$$

Below we write $W_1=W(x_1)$, $\v_1=\v(x_1)$, $I_1=I_\calB(x_1)$, $I_{1;2;-}=I{\{x_1\in\calB \mbox{ and }  x_2\in\calB\mbox{ and } x_3\notin\calB\}}$ and so forth. Then the required extremal coefficient is 
\begin{eqnarray*}
\theta_3&=&\E\left\{\max(W_1,W_2,W_3)\right\} \\
&=& {\sqrt{2\pi}\over \pr(x_1\in\calB)}\E\left\{\max\left(0,\v_1I_1,\v_2I_2,\v_3I_3\right)\right\}\\
&=&{\sqrt{2\pi}\over \pr(x_1\in\calB)}\bigg[\E\left\{\max\left(0,\v_1,\v_2,\v_3\right)I_{1;2;3}\right\} + \E\left\{\max\left(0,\v_1,\v_2\right)I_{1;2;-}\right\}\\
& & +\E\left\{\max\left(0,\v_1,\v_3\right)I_{1;-;3}\right\} + \E\left\{\max\left(0,\v_2,\v_3\right)I_{-;2;3}\right\}\\
& & +\E\left\{\max\left(0,\v_1\right)I_{1;-;-}\right\} + \E\left\{\max\left(0,\v_2\right)I_{-;2;-}\right\} + \E\left\{\max\left(0,\v_3\right)I_{-;-;3}\right\}\bigg]\\
&=& \pr(x_2\in\calB,x_3\in\calB\mid x_1\in\calB) \sqrt{2\pi} \E\left\{\max\left(0,\v_1,\v_2,\v_3\right)\right\}  + \pr(x_2\in\calB,x_3\notin\calB\mid x_1\in\calB) \sqrt{2\pi} \E\left\{\max\left(0,\v_1,\v_2\right)\right\}\\
& & + \pr(x_2\notin\calB,x_3\in\calB\mid x_1\in\calB) \sqrt{2\pi} \E\left\{\max\left(0,\v_1,\v_3\right)\right\} + \pr(x_1\notin\calB,x_3\in\calB\mid x_2\in\calB) \sqrt{2\pi} \E\left\{\max\left(0,\v_2,\v_3\right)\right\}\\
& & + \pr(x_2\notin\calB,x_3\notin\calB\mid x_1\in\calB)  + \pr(x_1\notin\calB,x_3\notin\calB\mid x_2\in\calB) + \pr(x_1\notin\calB,x_2\notin\calB\mid x_3\in\calB).
\end{eqnarray*}
The expression $\sqrt{2\pi} \E\left\{\max\left(0,\v_1,\v_2,\v_3\right)\right\}$ above is merely the trivariate extremal coefficient for the Schlather model without random sets, which can be evaluated quickly and accurately by simulation, whereas $\sqrt{2\pi} \E\left\{\max\left(0,\v_i,\v_j\right)\right\}$ is the bivariate extremal coefficient between station $i$ and station $j$, which can be computed analytically with the exponent measure $V_{i;j}(1,1)$.

The probabilities above correspond to the normalized expected volumes of overlap of three sets $\calB$ centered at $x_1$, $x_2$ and $x_3$. For example,  
$$\pr(x_2\in\calB,x_3\in\calB\mid x_1\in\calB)={\E\{|\calB\cap \{\calB+(x_2-x_1)\} \cap \{\calB+(x_3-x_1)\}|\}/ \E(|\calB|)},$$
$$\pr(x_2\in\calB,x_3\notin\calB\mid x_1\in\calB)={\E[|\calB\cap \{\calB+(x_2-x_1)\} \cap \{\calB+(x_3-x_1)\}^c|]/ \E(|\calB|)}.$$

For given radius $R$, duration $D$ and velocity $V$, the random set is fixed and the volume of overlap can be calculated analytically. Simulation can then be used to compute the expectation of such random quantities.

The same approach could be used to compute extremal coefficients at a higher order $N$, at the price of needing to compute by hand all the areas of overlap between $N$ discs with same radius.
\end{small}

\end{appendix}



\newpage

\end{document}